# Group Velocity Control in the four level system through interacting dark resonances

Kh. Saaidi[1], S. D. Seydi [2], S.W. Rabiei [3]

*Faculty of Science, Department of Physics, University of Kurdistan, Pasdaran Ave., Sanandaj, Iran*

**Abstract**

We investigate dispersion and the absorption properties of a weak probe field in a four level atomic system through interacting dark resonances. We show that two narrow peaks appears in the optical spectra due to the presence of interacting dark resonances. This two narrow peaks leads to the superluminal light propagation with strong absorption. However, if a weak incoherent pump field is exerted to the probe transition, then the peaks structure can be changed such that two absorption peaks transformed to two gain peaks. We show that both subluminal and superluminal light propagation occur in this system, where controlled by the incoherent pumping strength.

**Keywords**: Group velocity; Susceptibility; Absorption; Dispersion

[1]E-mail: ksaaidi@uok.ac.ir
[2]E-mail: sphysic84@yahoo.com
[3]E-mail: w.rabiei@yahoo.com

# 1 Introduction

A main interest in laser driven atomic media is the study of their coherence properties. Coherence effects as electromagnetically induced transparency (EIT) [1, 2] is example where the optical properties of an atomic medium is influenced by coherent fields. The propagation of a light pulse through a dispersive medium has been extensively investigated in the field of quantum coherence and interference [3, 4]. There are two kinds of velocities assosiated with a light pulse; a group velocity and phase velocity. The pulse as a whole travels at a speed known as the group velocity. In Ref. [5], Lord Rayleigh discussed that a pulse of light travels at the group velocity instead of the phase velocity inside a medium and further developed the theory on transparency and anomalous dispersion. It is well know that the propagation of a light pulse can be slowed down, or it can become faster than c (light propagation velocity in vacuum), or can even become negative in a transparent medium [3, 4, 6]. The superluminal light propagation has group velocity greater than of the light in vacuum, while the subluminal light propagation has group velocity down than of the light in vacuum. The superluminal light propagation in a dispersive medium originates from the interference of the different frequency components of the light pulse. Thus the coherence of such components has a major role in superluminal light propagation. The light propagation by reducing the coherence of the light pulse changes from superluminal to subluminal. That superluminal light propagation communications and travel refer to the propagation of information or matter faster than the speed of light. Under the special theory of relativity, a particle with subluminal velocity needs infinite energy to accelerate to the speed of light, although special relativity does not forbid the existence of particles that travel faster than light at all times.

The theory of superluminal propagation has actively been developed since $1990'\text{s}$ [7, 8]. Recently, slow or even stopped group velocities and superluminal propagation of light waves have been demonstrated in specific classes of atomic and solid state materials [9].

Here, we demonstrated that the incoherent pump field has an important role in the controlling the group velocity of the light pulse in a dispersion medium. In this paper, we investigate probe pulse propagation via a system which exhibits interacting dark resonances. Our studied system is recognized, e.g., in mercury. We see that the medium susceptibility in dependence on the probe field detuning display high contrast structures characteristic of interacting dark states. Here, our main purpose is the response of the atomic medium to the probe field. We illustrate where the presence two perturbing fields with Rabi frequency $\Omega_{s_1}$ and $\Omega_{s_2}$ on transition $|4\rangle \leftrightarrow |3\rangle$, leads in dark resonances. Hence, we find that superluminal and subluminal light propagation occurs around the two absorption spikes so-called dark resonances. Where as we apply two coherent coupling fields on transition $|3\rangle \leftrightarrow |4\rangle$, we can not eliminate the time parameter in evolution equation of system. So that we must solve this model by using the Floquet method. Also these two coherent coupling fields, which is applied on $|3\rangle \leftrightarrow |4\rangle$, cause two absorption (gain) spikes. Then we have more region for superluminal and subluminal light.

The scheme of this paper is as following. In Sec. 2, we consider theoretical analysis on a four-level atomic system and we obtain the evolution equation for density matrix. In Sec. 3, we solve the time-dependent equations of density matrix by Floquet method. In Sec. 4, we interpret the obtained result and in Sec. 5, we analysis and discuss the numerical result of the paper.

## 2 Theoretical Analysis

We consider a four-level atomic systems as shown in figure 1. A strong coherent pump field of frequency $w_c$ and amplitude $\vec{E}_c$ with the Rabi frequency $\Omega_c = \vec{E}_c \cdot \vec{d}_{23}/\hbar$ drives the $|2\rangle \leftrightarrow |3\rangle$ transition, while a weak tunable probe field of frequency $w_p$ and amplitude $\vec{E}_p$ with Rabi frequency $\Omega_p = \vec{E}_p \cdot \vec{d}_{12}/\hbar$ applies to the $|1\rangle \leftrightarrow |2\rangle$ transition. Two weak coupling coherent fields with frequency $w_{s_1}, w_{s_2}$ and Rabi frequency $\Omega_{s_1} = \vec{E}_{s_1} \cdot \vec{d}_{43}/\hbar$ and $\Omega_{s_2} = \vec{E}_{s_2} \cdot \vec{d}_{43}/\hbar$ is applied to transition $|4\rangle \leftrightarrow |3\rangle$. Here $\vec{d}_{ij}$ are the atomic dipole moments. We include spontaneous decay with rates $\gamma_{32}, \gamma_{34}, \gamma_{21}$ and $\gamma_{41}$, respectively, on the dipole-allowed transition.

One can defined, $\Delta_c = w_c - w_{32}, \Delta_p = w_p - w_{21}, \Delta_{s_1} = w_{s_1} - w_{34}$, and $\Delta_{s_2} = w_{s_2} - w_{34}$. These quantities are the detuning of the drive field, the probe field and two the coherent perturbation fields, respectively.

Finally, an incoherent driving field with pump stretch $r$ is applied on the probe transition $|1\rangle \leftrightarrow |2\rangle$, which the population of the level $|1\rangle$ can be pump to the level $|2\rangle$.

The Hamiltonian in dipole and interaction picture and the rotating wave approximation is given by:

$$\begin{aligned} H &= -\hbar\Omega_p e^{i\Delta_{12}t}|1\rangle\langle 2| - \hbar\Omega_c e^{i\Delta_c t}|2\rangle\langle 3| - \hbar(\Omega_{s_1} e^{i\Delta_{s_1}t} \\ &+ \Omega_{s_2} e^{i\Delta_{s_2}t})|4\rangle\langle 3| + c.c., \end{aligned} \quad (1)$$

now, we debut the master equation for the atomic density matrix. The unitary evolution equations is given by the Von-Neumann equation and the

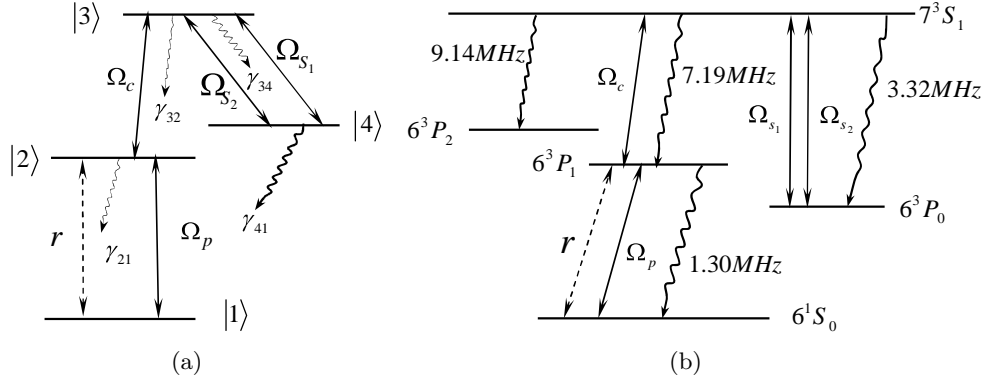

Figure 1: (a) Energy design of the four level system displaying. (b) A possible perception of the design in mercury.

spontaneous decay rate [1]. In density matrix equations of motion, in following by the rotating wave approximations for equations, time-dependent eliminate on $\Omega_{s_1}$ and only time-dependent appear aside $\Omega_{s_2}$. Therefore, the evolution equations of density matrix, in the rotating wave approximations are as

$$\dot{\rho_{11}} = \gamma_{21}\rho_{22} + r(\rho_{22} - \rho_{11}) + i\Omega_p \rho_{21} - i\Omega_p^* \rho_{12} + \gamma_{41}\rho_{44}, \tag{2}$$

$$\dot{\rho_{22}} = \gamma_{32}\rho_{33} - \gamma_{21}\rho_{22} + r(\rho_{11} - \rho_{22}) + i\Omega_p^* \rho_{12} + i\Omega_c \rho_{32} - i\Omega_c^* \rho_{23} - i\Omega_p \rho_{21}, \tag{3}$$

$$\dot{\rho_{44}} = \gamma_{34}\rho_{33} + i(\Omega_{s_1} + \Omega_{s_2} e^{i(\Delta t + \phi)})\rho_{34} - i(\Omega_{s_1} + \Omega_{s_2} e^{-i(\Delta t + \phi)})\rho_{43} - \gamma_{41}\rho_{44}, \tag{4}$$

$$\dot{\rho_{21}} = i\Delta_p \rho_{21} + i\Omega_p^* (\rho_{11} - \rho_{22}) + i\Omega_c \rho_{31} - \frac{1}{2}(\gamma_{21} + 2r)\rho_{21}, \tag{5}$$

$$\dot{\rho_{31}} = i(\Delta_p + \Delta_c)\rho_{31} + i(\Omega_{s_1} + \Omega_{s_2} e^{-i(\Delta t + \phi)})\rho_{41} + i\Omega_c^* \rho_{21} - i\Omega_p^* \rho_{32} - \frac{1}{2}(\gamma_{34} + \gamma_{32} + r)\rho_{31}, \tag{6}$$

$$\dot{\rho_{32}} = i\Delta_c \rho_{32} + i\Omega_c^* (\rho_{22} - \rho_{33}) + i(\Omega_{s_1} + \Omega_{s_2} e^{-i(\Delta t + \phi)})\rho_{42} - i\Omega_p \rho_{31} - \frac{1}{2}(\gamma_{34} + \gamma_{32} + \gamma_{21} + r)\rho_{32}, \tag{7}$$

$$\dot{\rho_{34}} = i\Delta_{s_1}\rho_{34} + i\Omega_c^* \rho_{24} + i(\Omega_{s_1} + \Omega_{s_2} e^{-i(\Delta t + \phi)})(\rho_{44} - \rho_{33}) - \frac{1}{2}(\gamma_{34} + \gamma_{32} + \gamma_{41})\rho_{34}, \tag{8}$$

$$\dot{\rho_{41}} = i(\Delta_p + \Delta_c - \Delta_{s_1})\rho_{41} + i(\Omega_{s_1} + \Omega_{s_2} e^{i(\Delta t + \phi)})\rho_{31} - i\Omega_p^* \rho_{42},$$

$$-\frac{1}{2}(\gamma_{41}+r)\rho_{41}, \tag{9}$$

$$\dot{\rho_{42}} = i(\Delta_c - \Delta_{s_1})\rho_{42} + i(\Omega_{s_1}+\Omega_{s_2}e^{i(\Delta t+\phi)})\rho_{32} - i\Omega_c^*\rho_{43} - i\Omega_p\rho_{41}$$
$$-\frac{1}{2}(\gamma_{21}+\gamma_{41}+r)\rho_{42}, \tag{10}$$

$$\rho_{11} + \rho_{22} + \rho_{33} + \rho_{44} = 1. \tag{11}$$

In this equations, $\Gamma_{ij}=(\gamma_i+\gamma_j)/2$ are the damping rates of the coherence with $\gamma_i$ begin the total decay rate out of state $|i\rangle$, $(i,j \in 1,2,...,4)$

## 3  Time-Dependent Solution

We now evaluate the equations of motion equations $(2-11)$ for the general case of time-dependent coefficients. This set of equations can be solved by method so-called the Floquet method, by writing in the matrix from as

$$\frac{\partial R}{\partial t} + \Sigma = MR, \tag{12}$$

where

$$R = \big(\rho_{11},\rho_{12},\rho_{13},\rho_{14},\rho_{21},\rho_{22},\rho_{23},\rho_{24},\rho_{31},\rho_{32},\rho_{34},\rho_{41},\rho_{42},\rho_{43},\rho_{44}\big)^T, \tag{13}$$

and

$$\Sigma = \big(0,0,0,0,0,-\gamma_{32},-i\Omega_c,0,0,i\Omega_c,i(\Omega_{s_1}+\Omega_{s_2}e^{-i(\Delta t+\phi)}),0,0,$$
$$- i(\Omega_{s_1}+\Omega_{s_2}e^{i(\Delta t+\phi)}),-\gamma_{34}\big)^T. \tag{14}$$

Here R, $\sum$ are vectors containing the density matrix elements and a vector independent of the density matrix elements, respectively. Where $\sum$ stems from eliminating of $\rho_{33}$. The matrix M follows from equations $(2-11)$ and $\sum$ can be separated into terms with different time dependence, as

$$\Sigma = \Sigma_0 + \Sigma_1 e^{-i(\Delta t+\phi)}\Omega_{s_2} + \Sigma_{-1}e^{i(\Delta t+\phi)}\Omega_{s_2}, \tag{15}$$
$$M = M_0 + M_1 e^{-i(\Delta t+\phi)}\Omega_{s_2} + M_{-1}e^{i(\Delta t+\phi)}\Omega_{s_2}, \tag{16}$$

where $\Sigma_j$ and $M_j (j \in 0, \pm 1)$ are time-independent.
It is major to consider that the time dependence of equations $(2-11)$ only stems from the parameter $\Delta$. According to Floquet method [10, 11], the solution R therefore has only contributions oscillating at harmonics of the detuning $\Delta$.

We have to solve equation (12) including the explicit time dependence. So we expand R in terms of $\Omega_{s_2}$, [12]

$$R = \sum_{n=0}^{\infty} R_n \Omega_{s_2}^n. \tag{17}$$

Thus we make an ansatz for the solution and write $R_n$ in a Fourier series,

$$R_n = \sum_{m=-\infty}^{\infty} R_n^{(m)} e^{-im(\Delta t + \phi)}. \tag{18}$$

First by using equation (18) in equation (17) and hence by inserting equation (17) and (16) and (15) in equation (12) and equating the coefficients oscillating at different harmonics of $\Delta$. One can derive a hierarchy of time independent equations for the coefficients $R_n^m$. Up to order $\mathcal{O}[\Omega_{s_2}^5]$ we find:

$$\begin{align}
R_0^{(0)} &= M_0^{-1} \Sigma_0, \tag{19} \\
R_1^{(\pm 1)} &= (M_0 \pm i\Delta I)^{-1} (\Sigma_{\pm 1} - M_{\pm 1} R_0^{(0)}), \tag{20} \\
R_2^{(0)} &= -M_0^{-1} (M_{-1} R_1^{(1)} + M_1 R_1^{(-1)}), \tag{21} \\
R_2^{(\pm 2)} &= -(M_0 \pm 2i\Delta I)^{-1} M_{\pm 1} R_1^{(\pm 1)}, \tag{22} \\
R_3^{(\pm 1)} &= -(M_0 \pm i\Delta I)^{-1} (M_{\pm 1} R_2^{(0)} + M_{\mp 1} R_2^{(\pm 2)}), \tag{23} \\
R_3^{(\pm 3)} &= -(M_0 \pm 3i\Delta I)^{-1} M_{\pm 1} R_2^{(\pm 2)}, \tag{24} \\
R_4^{(0)} &= -M_0^{-1} (M_1 R_3^{(-1)} + M_{-1} R_3^{(1)}), \tag{25} \\
R_4^{(\pm 2)} &= -(M_0 \pm 2i\Delta I)^{-1} (M_{\pm 1} R_3^{(\pm 1)} + M_{\mp 1} R_3^{(\pm 3)}), \tag{26} \\
R_4^{(\pm 4)} &= -(M_0 \pm 4i\Delta I)^{-1} M_{\pm 1} R_3^{(\pm 3)}, \tag{27} \\
R_5^{(\pm 1)} &= -(M_0 \pm i\Delta I)^{-1} (M_{\pm 1} R_4^{(0)} + M_{\mp 1} R_4^{(\pm 2)}), \tag{28} \\
R_5^{(\pm 3)} &= -(M_0 \pm 3i\Delta I)^{-1} (M_{\pm 1} R_4^{(\pm 2)} + M_{\mp 1} R_4^{(\pm 4)}), \tag{29} \\
R_5^{(\pm 5)} &= -(M_0 \pm 5i\Delta I)^{-1} M_{\pm 1} R_4^{(\pm 4)}, \tag{30}
\end{align}$$

where $I$ is the unit matrix and $M_0$ is time-independent, all other $R_n^{(m)}$ up to the this order vanish. In general we obtain that

$$R = \sum_{n=0}^{\infty} \sum_{m=-n,-n+2,\ldots}^{n} R_n^{(m)} \Omega_{s_2}^n e^{-im(\Delta t + \phi)}. \tag{31}$$

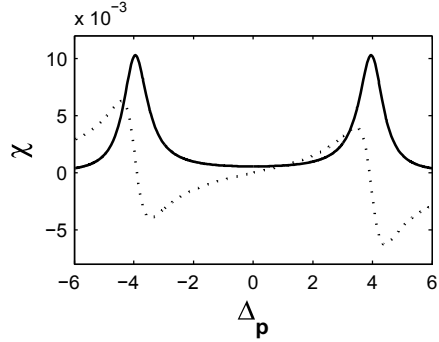

Figure 2: The Real (dashed) and imaginary (solid) parts of the $\chi$ as a function of the probe detuning $\Delta_p$ for the parameters $\Omega_p = 0.01, \Omega_c = 4, \Omega_{s_1} = 0, \Omega_{s_2} = 0, \Delta_{s_1} = 0.2, \Delta_{s_2} = -0.2, r = 0, \gamma_{21} = 0.14, \gamma_{34} = 1, \gamma_{41} = 0.01, \gamma_{32} = 0.79$.

## 4   Physical Interpretation

To definition of the various coefficients, we study the effect of the different parts of the solution on the probe field. Since we write the expansion series for the relevant probe field coherence in the schrodinger picture $\rho_{21}$ in down using the explicit transformation relation connecting the schrodinger picture with our interaction picture. We obtain

$$\rho_{21}^s = \rho_{21} e^{-i\omega_p t}. \tag{32}$$

Where $\rho_{21}$ is the density matrix in the interaction picture, and gives as component of the solution for R we find

$$\rho_{21} = \sum_{n=0}^{\infty} \sum_{m=-n,-n+2}^{n} [R_n^{(m)}]_5 \Omega_{s_2}^n e^{-i\omega_p t} e^{-im(\Delta t + \phi)}. \tag{33}$$

In relation (33), $[R_n^{(m)}]_5$ refers to the five component of vector $R_n^{(m)}$. It is seen that $\chi$ oscillate with probe frequency, $\omega_p$, only for $m = 0$ and then we see that the contribution proportional to three terms $R_0^{(0)}, R_2^{(0)}$ and $R_4^{(0)}$ oscillate with the probe frequency. Therefore, only this three terms contribute in the susceptibility. Therefore the response of the atomic system to the applied field, $\chi$, is obtained as

$$\chi(\omega_p) = \frac{2Nd_{12}}{\epsilon_0 E_p} [[R_0^{(0)}]_5 + [R_2^{(0)}]_5 \times \Omega_{s_2}^2 + [R_4^{(0)}]_5 \times \Omega_{s_2}^4]. \tag{34}$$

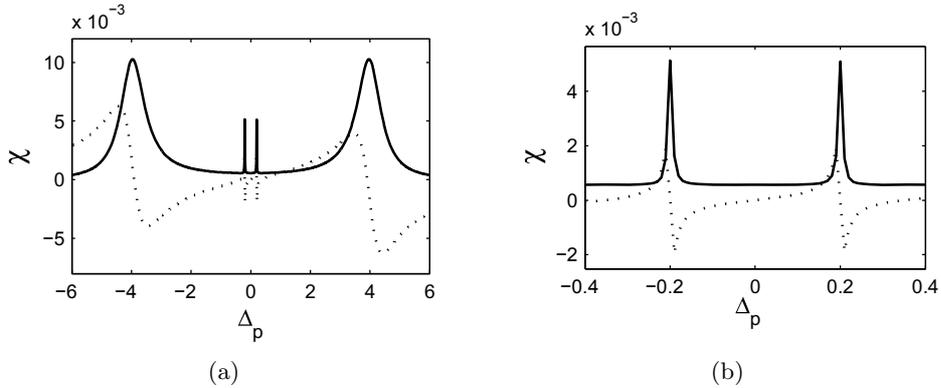

(a)                            (b)

Figure 3: (a)The Real (dashed) and imaginary (solid) parts of the $\chi$ as a function of the probe detuning $\Delta_p$ for the parameters $\Omega_p = 0.01, \Omega_c = 4, \Omega_{s_1} = \Omega_{s_2} = 0.2, \Delta_{s1} = 0.2, \Delta_{s2} = -0.2, r = 0.0, \gamma_{21} = 0.14, \gamma_{34} = 1, \gamma_{41} = 0.01, \gamma_{32} = 0.79$. (b) is a closeup on the central part of (a).

Here N is the atom number density in the medium, $d_{12}$ is the probe transition dipole moment. For the realistic example, we consider $\frac{2Nd_{12}}{\epsilon_0 E_p} \cong 1$. Indeed, we see that only three terms with $m = 0$ oscillate by probe frequency, and these three terms contributes in the susceptibility, $\chi$. We indicate $\chi$ as $\chi = \chi' + i\chi''$, where the real and imaginary parts of $\chi$ correspond to the dispersion and the absorption of the weak probe field, respectively. The group velocity $V_g$ of the weak probe field is then given by [13, 14, 15, 16]

$$V_g = \frac{c}{1 + 2\pi\chi'(\omega_p) + 2\pi\omega_p \frac{\partial \chi'(\omega_p)}{\partial \omega_p}} = \frac{c}{n_g}, \qquad (35)$$

where c is the speed of light in the vacuum, and shows that, $n_g$ is the group index. Equation (35), the group velocity can be reduced via a steep positive dispersion. On the other hand, strong negative dispersion can lead to an increase in the group velocity and even to a negative group velocity.

## 5   Numerical results

In figure 2, we show the real and imaginary part of the $\chi$ which correspond to the dispersive and absorptive properties of the medium, respectively, in terms of the probe field detuning $\Delta_p = \Delta_{21}$. In this figure, the two perturbing laser field are turn off, $\Omega_{s_1} = \Omega_{s_2} = 0$. The common parameters are $\Omega_p = 0.01, \Omega_c = 4, \Delta_{s_1} = 0.2, \Delta_{s_2} = -0.2, r = 0, \gamma_{21} = 0.14, \gamma_{34} = 1, \gamma_{41} = 0.01, \gamma_{32} = 0.79$.

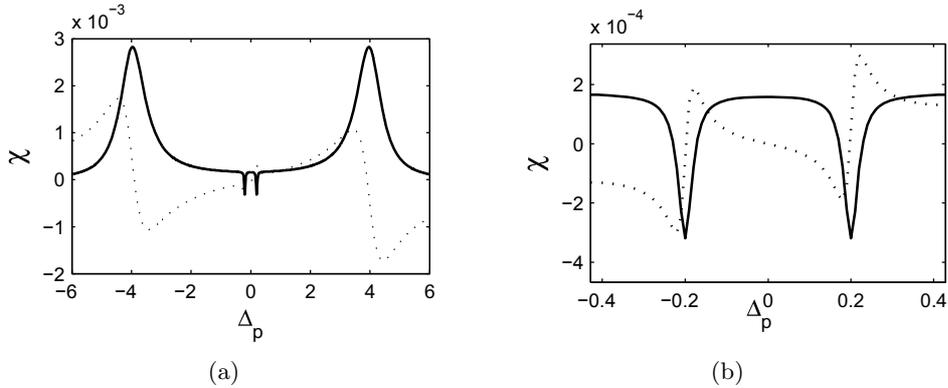

(a)  (b)

Figure 4: (a)the Real (dashed) and imaginary (solid) parts of the $\chi$ as a function of the probe detuning $\Delta_p$ for the parameters $\Omega_p = 0.01, \Omega_c = 4, \Omega_{s_1} = \Omega_{s_2} = 0.2, \Delta_{s_1} = 0.2, \Delta_{s_2} = -0.2, r = 0.03, \gamma_{21} = 0.14, \gamma_{34} = 1, \gamma_{41} = 0.01, \gamma_{32} = 0.79$. (b) is a closeup on the central part of (a).

We add a weak decay rate $\gamma_{41}$. For this case, in the steady state, all population is trapped in $|1\rangle$. The driving field with Rabi frequency $\Omega_c$ leads to double absorption peaks at $\Delta_p = \pm 4$. It is clearly seen that, this system has not any absorption around zero detuning. So that this model has partially electromagnetically induced transparency (EIT), i.e., a dip in the absorption at zero detuning. In this figure, the slope of dispersion (the real part of the $\chi$ ) in the region of reduced absorption is positive, thus subluminal light propagation occurs around zero detuning with reduced absorption as it is common for EIT. In figure 3(a), we see two narrow absorption spikes around zero detuning. This results arises from adding two weak perturbing fields with Rabi frequency $\Omega_{s_1} = \Omega_{s_2} = 0.2$. The results in this figure are identical to figure 2 except for two narrow absorption spikes around zero detuning. The slope of the real part of the $\chi$ around $\Delta_p = \pm 0.2$ is negative and corresponding to the superluminal light propagation albeit with high absorption. In this figure, the slope of the dispersion is positive between two closely spaced absorption line at $\Delta_p = 0.2$ and $\Delta_p = -0.2$. This leads to a slowing down of the group velocity, and then it can be made the subluminal light propagation.

Now, we investigate the effect of the weak incoherent pumping field on the probe transition $|2\rangle - |1\rangle$. Furthermore, in figure 4, we see the dispersion and the absorption properties of the probe field versus probe detuning for incoherent pumping rates. In figure 4 by presence of the incoherenct pump field rate, i.e. r=0.03, two absorption spikes at $\Delta_p = \pm 0.2$ transforms two gain spikes and the superluminal light propagation found in figure 3 at $\Delta_p = \pm 0.2$ switches to subluminal propagation. Where this results found in part $b$ of figure 4. In figure 4(b), the slope of the dispersion is negative between two gain line at $\Delta_p = \pm 0.2$. This leads to the superluminal light propagation.
In figure 5, we plot the imaginary parts of the $\chi$ as a function of the probe detuning $\Delta_p$ for different values of incoherent pump field for $r = 0$, $r = 0.009$. Figure 5 show that by increasing the incoherent pumping rate, the absorption peak is decreased and finally the two absorption spikes at around zero detuning become two gain spikes for $r = 0.03$. In figure 6 we plot the real parts of the $\chi$ as a function of the probe detuning $\Delta_p$ for different values of incoherent pump field for $r = 0$, $r = 0.009$. Figure 6 shows the corresponding results. In this figure, we show that by increasing the incoherent pumping rate, the slope of the dispersion ( real part of the susceptibility ) is changed and finally, the slope of the dispersion of negative found in figure 3 at $\delta_p = \pm 0.2$ switches to positive in figure 4. Therefore we see that the superluminal light propagation in figures 3 for $r = 0$ at $\Delta_p = \pm 0.2$ switches to subluminal light propagation in figure 4 for $r = 0.03$.

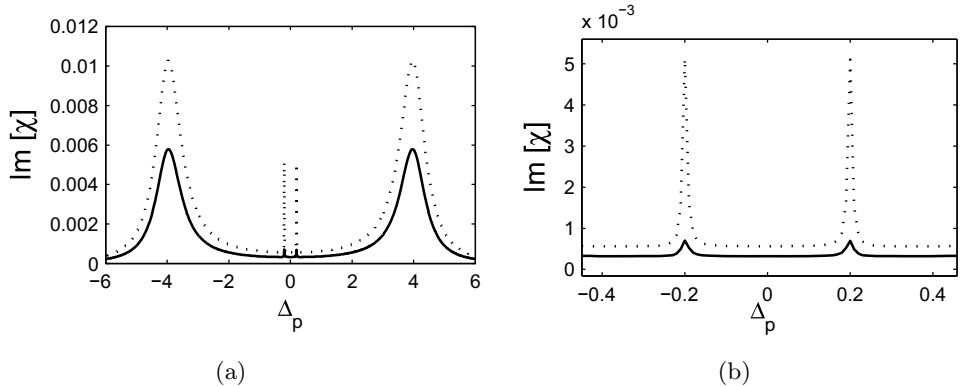

Figure 5: (a)The imaginary parts of the $\chi$ as a function of the probe detuning $\Delta_p$ for the parameters $r = 0$ (dashed), $r = 0.009$ (solid ). The other parameters are the same as in figure 3(a). (b) is a closeup on the central part of (a).

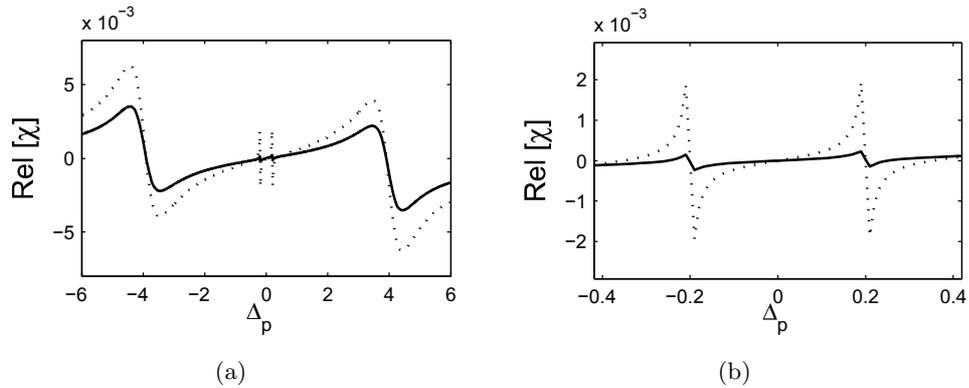

Figure 6: (a)The Real parts of the $\chi$ as a function of the probe detuning $\Delta_p$ for the parameters $r = 0$ (dashed), $r = 0.009$ (solid ). The other parameters are the same as in figure 3(a). (b) is a closeup on the central part of (a).

## 6 Conclusion

In this paper we have investigated the dispersive and absorptive properties of a four level system. We exhibits two absorption spikes around zero detuning. A weak probe field, in absence an incoherent pump field experiences superluminal propagation with strong absorption, while in present this incoherent pump field on the probe transition, the superluminal light propagation changes to subluminal light propagation. Also, we show that by applying a weak incoherent pumping field on the probe transition, there are both superluminal and subluminal light around two gain spikes, where the control via the incoherent pumping field strength is possible.